\documentclass[times, 10pt,twocolumn]{article} 
\usepackage{latex8}
\usepackage{times}

\usepackage{rotating}

\begin{document}

\title{Attributes for Causal Inference in Longitudinal Observational Databases}

\author{Jenna Reps, Jonathan M. Garibaldi, \\ Uwe Aickelin and Daniele Soria\\
IMA, School of Computer Science \\ The University of Nottingham \\ Nottingham NG8 1BB \\ 
\{jzr,jmg,uxa,dqs\}@cs.nott.ac.uk\\
\and
Jack E. Gibson and Richard B. Hubbard\\
Clinical Sciences Building \\
Nottingham City Hospital\\
Nottingham, NG5 1PB\\
\{jack.gibson,\\richard.hubbard\}@nottingham.ac.uk\\
}

\maketitle
\thispagestyle{empty}

\begin{abstract}
The pharmaceutical industry is plagued by the problem of side effects that can occur anytime a prescribed medication is ingested. There has been a recent interest in using the vast quantities of medical data available in longitudinal observational databases to identify causal relationships between drugs and medical events. Unfortunately the majority of existing post marketing surveillance algorithms measure how dependant or associated an event is on the presence of a drug rather than measuring causality.  In this paper we investigate potential attributes that can be used in causal inference to identify side effects based on the Bradford-Hill causality criteria.  Potential attributes are developed by considering five of the causality criteria and feature selection is applied to identify the most suitable of these attributes for detecting side effects. We found that attributes based on the specificity criterion may improve side effect signalling algorithms but the experiment and dosage criteria attributes investigated in this paper did not offer sufficient additional information. 

\end{abstract}

\Section{Introduction}

The majority of the population will require medication at some point during their lifetime and fortunately the medication will often increase their standard of living.  Occasionally and often unpredictably a patient may experience a negative side effect as a consequence of ingesting a drug.  When a drug is known to have caused a negative side effect, the side effect is referred to as an adverse drug reaction (ADR).  By the previous definition of an ADR it is clear that an ADR represents a causal relation between a drug and medical event.  

The majority of existing algorithms for signalling ADRs identify medical events that are associated to a specific drug but association does not imply causation, as for example, a drug and medical event may be associated if the drug is prescribed to treat an illness that always precedes the medical event.  As a consequence, the existing algorithms generate many false positive signals and are only considered ADR filters rather than definitive detectors of ADRs as the signals they generate still need to be confirmed by more rigorous investigations \cite{Almenoff2005}.  If an algorithm was developed that could definitively detect ADRs then this algorithm is likely to signal ADRs at a quicker rate due to no longer requiring the rigorous investigations and would enable the investigation of all drug and medical event pairs rather than just the ones that seem to have the greatest association.

One possible way of developing an algorithm that is capable of classifying each pair of drug and medical event as either an ADR (the drug causes the medical event), an indicator (the medical event causes the drug) or noise (no clear causal relationship between the drug and medical event) is to train a learning algorithm with suitable attributes.  The attributes need to be selected so that they give information about causal relations.  In this paper we have decided to develop and investigate attributes derived from the Bradford-Hill causality criteria \cite{Hill1965} and find the most suitable attributes by feature selection.  The Bradford Hill causality criteria are a group of nine factors that are often considered when determining causality in epidemiological studies.  This is the first time attributes to determine causation in Longitudinal Observational Databases (LODs) have been investigated. The purpose of the paper is to firstly determine if including attributes that cover previously unconsidered Bradford Hill causality criteria could be advantageous for algorithms that signal ADR in LODs and secondly to predict how useful each attribute is. The results of this paper will help any future ADR classifying algorithm.  

The continuation of this paper is are follows. In the next section the existing methods and Bradford Hill causality criteria are discussed.  This is followed by the methodology section detailing the attributes of interest for each of the criterion chosen and the feature selection method applied.  Section \ref{results} is the results of the feature selection and section \ref{diss} discusses the implications of our results.  The paper concludes with section \ref{con} and possible future work is suggested.

\section{Background}

Existing methods of signalling ADR in healthcare databases have generally been developed and applied to Spontaneous Reporting System (SRS) databases, but new methods are being developed for LODs.  The SRS databases contain voluntary records of suspected ADRs, for example, if a patient thinks they have experienced an ADR they can report this and the report will be added into an SRS database.  These databases often contain a vast quantity of reports and have been used successfully in the past to identify some ADRs. Unfortunately the SRS databases are limited in the ADRs they can detect due to some ADRs being difficult to spot and are therefore unlikely to be reported and the reports often contain missing, incorrect or duplicated data \cite{Almenoff2005}. The SRS databases are known to suffer from underreporting \cite{Hazell2006} as people may not identify an ADR or may not bother to report one and this can cause a time lag before an ADR is signalled.  The ADR signalling algorithms developed for SRS databases are often referred to as `disproportionality methods' as they estimate the background rate that a specific medical event occurs by considering how often it is reported with any drug and compare this with the how often it is reported with the drug of interest \cite{Puijenbroek2002}.  The disproportionality methods signal medical events that are highly associated to a specific drug, but these signalled medical events then require further investigation to determine if the association is causal or not.    

The LOD algorithms generally determine how dependant the occurrence of a medical event is on the occurrence of the drug by contrasting the observed number of times the medical event occurs after the drug with how often you would expect the medical event to occur after the drug under the assumption they are independent.  The methods deviate in how they calculate the expected number of times a medical event will occur after the drug, some use a cohort approach \cite{jin2006} \cite{jin2010}, while others use a mixture of the case-series and the `disproportionality' approaches \cite{noren2010}.  These algorithms also tend to consider a temporal relation by filtering medical events that are considered to have caused the drug to be taken, or medical events that are likely to be `therapeutic failures' (a result of the drug not working) \cite{Schuemie2011}.  Other ADR signalling algorithms have used patient time to calculate an expectation  \cite{Schuemie2011} or applied a sequential case-series approach  \cite{Hocine2009}.

The Bradford Hill causality criteria were developed as a way to distinguish between association and causation.  The nine factors of interest are summarised below.  
\begin{itemize}\addtolength{\itemsep}{-0.7\baselineskip}
\item Association Strength - how strong the association is. 
\item Temporality - the direction of the association, does the medical event often appear after the drug or before it?
\item Specificity - how specific the relationship is. 
\item Experimentation - when the drug stops does the medical event stop, when it restarts does the medical event also restart? 
\item Dosage - is there correlation between dosage and experiencing the medical event?
\item Analogy - does a similar drug have a similar side effect?
\item Coherence - does the association make sense?
\item Plausibility - is the association possible?
\item Consistency - is there evidence of the association in different databases?
\end{itemize}
\begin{table}[ht]
\caption{The Bradford-Hill causality criteria and whether they are covered by existing methods.}
\label{tab:bhc}
\begin{tabular}{ccc}
Criterion &  SRS methods & LOD methods  \\ \hline
Association Strength & Yes & Yes \\
Temporality &  Yes & Yes \\
Specificity  & No & No \\
Experimentation &  No & No\\
Dosage  & No & No \\
Analogy  & No & No \\
Coherence  & Indirectly & No \\ 
Plausibility  & No & No \\
Consistency & No & No \\
\end{tabular}
\end{table}
Table \ref{tab:bhc} summarises the criteria covered by the existing methods.  It is clear that the existing methods all cover association strength and temporality, as patients are in general only likely to report a potential ADR to an SRS database if it is newly occurring and the LOD algorithms include temporal filters.  The SRS algorithms also indirectly cover plausibility as the people making the report will natural filter any medical event that occurs after the drug is prescribed that is not plausible as an ADR.  The remaining criteria are not covered by the existing methods. However, attributes that cover association strength, temporality, specificity, experimentation and dosage are developed in the next section. Future algorithms integrating these previously unconsidered attributes for specificity, experimentation and dosage may offer new in sight for ADR detection.       

\Section{Methodology}
\subsection{Step 1: Generating Attributes}
\subsubsection{Strength of Association}	
The strength of association measures how dependant the medical event is on the presence of the drug but does not consider confounding effects.  The standard epidemiology and pharmacovigilance methods for calculating association strength often make use of a 2x2 contingency table, see Table \ref{fig:conf_tab}, that compares the frequency that a medical event is observed in a group of patients prescribed the drug of interest and a group of patients prescribed any other drug.
\begin{table*}
\caption{A 2x2 contingency table frequently used in pharmacovigilance studies.}
\label{fig:conf_tab}
\centering
\begin{tabular}{c|ccc}
& Event of Interest (X=1) & Other Event (X=0) & Totals \\ \hline
Drug of Interest (Y=1) & a & b & a+b \\
Other Drug (Y=0) & c & d & c+d \\
Totals & a+c & b+d & a+b+c+d \\
\end{tabular}
\end{table*}
Three common measures of association strength are the risk difference, risk ratio and odds ratio.  The risk of having medical event  in the month after the drug of interest is prescribed is then $a/(a+b)$ and the risk of having the event  in the month after any other drug is $c/(c+d)$. The risk difference investigates the difference whereas the risk ratio investigates the ratio between the risks of the two groups.  The odds ratio calculates the ratio between the odds that a medical event occurs in the drug of interest group ($a/b$) and the odds that a medical event occurs in the any other drug group ($c/d$).  The different association strength measures and their probabilistic interpretations are described in Table \ref{tab:ass_str}.
\begin{table*}
\centering
\caption{The different association strength measures and their probabilistic interpretations.}
\label{tab:ass_str}
\begin{tabular}{c|cc}
Association Measure & Probabilistic Interpretation & Calculation \\ \hline
Risk Difference (RD) & $P[X=1|Y=1]-P[X=1|Y=0]$ & $\frac{a}{a+b}-\frac{c}{c+d}$ \\
Risk Ratio (RR) & $P[X=1|Y=1]/P[X=1|Y=0]$ & $\frac{a}{a+b}/\frac{c}{c+d}$ \\
Odds Ratio (OD) & $\frac{P[X=1|Y=1]/P[X=0|Y=1]}{P[X=1|Y=0]/P[X=0|Y=0]}$ & $ad/bc$ \\ 
\end{tabular}
\end{table*}
\begin{table*}
\caption{Attribute Summary Table}
\label{tab:attributes}
\begin{tabular}{ccl}
Feature & Criterion & Description \\ \hline
RR, RD, OR & Strength & \parbox[t]{11.5cm}{The Risk Ratio, Risk Difference and Odds Ratio for all prescriptions.} \\
\parbox[t]{3cm}{RR$_{13d}$,RD$_{13d}$,OR$_{13d}$} & Strength & \parbox[t]{11.5cm}{The Risk Ratio, Risk Difference and Odds Ratio for drugs prescribed for the first time in 13 months.} \\
\parbox[t]{3cm}{RR$_{13BNF}$,RD$_{13BNF}$ ,OR$_{13BNF}$} & Strength & \parbox[t]{11.5cm}{The Risk Ratio, Risk Difference and Odds Ratio for drugs corresponding to a bnf that has not been prescribed in the last 13 months.} \\
IC$_{\Delta}$ & Strength & \parbox[t]{11.5cm}{The Information Component as calculated in \cite{noren2010}}\\
lowerIC$_{\Delta}$ & Strength & \parbox[t]{11.5cm}{The lower 95\% interval of the Information Component as calculated in \cite{noren2010}} \\
LEOPARD & Temporality & \parbox[t]{11.5cm}{Is $1$ if the drug is prescribed significantly more after the medical event than before and $0$ otherwise.}  \\
$OE_{filt1}$ & Temporality & \parbox[t]{11.5cm}{Calculates if the IC$_{\delta}$ is greater the month before the drug than the month after.}\\
$OE_{filt2}$ & Temporality & \parbox[t]{11.5cm}{Calculates if the IC$_{\delta}$ is greater on the day of prescription compared to the month after.}\\
ABratio Level 2 & Temporality & \parbox[t]{11.5cm}{How often the level 2 version of a Read Code is recorded after the prescription compared to before.} \\
ABratio Level 3 & Temporality & \parbox[t]{11.5cm}{How often the level 3 version of a Read Code is recorded after the prescription compared to before.} \\
\parbox[t]{3cm}{Age Standard Deviation} & Specificity & \parbox[t]{11.5cm}{Standard deviation of patient's age who experience medical event after drug divided by standard deviation of the ages for all the patients.}  \\
Gender Ratio & Specificity & \parbox[t]{11.5cm}{Male proportion of patients experiencing the medical event within 30 days of the drug divided by male proportion of patients prescribed the drug.} \\
Read Code Level & Specificity & \parbox[t]{11.5cm}{The specificity level of the Read Code.} \\
RR drug /  RR bnf  & Specificity &  \parbox[t]{11.5cm}{The RR of the drug divided by the RR of the drug's corresponding BNF.}  \\
Dosage Ratio & Dosage & \parbox[t]{11.5cm}{Average dosage of patients experiencing the medical event within 30 days of the drug divided by average dosage of patients prescribed the drug.} \\
High Low Ratio & Dosage & \parbox[t]{11.5cm}{Proportion of patients given the highest dosage that experience the medical event (within 30 days) divided by the proportion of patients given the lowest dosage that experience the medical event (within 30 days).} \\
Spearman's rank & Dosage & \parbox[t]{11.5cm}{The Spearman's rank correlation coefficient between the patient dosage and $\{0,1\}$ indicating if the patient experienced the medical event within 30 days.} \\
\parbox[t]{2.5cm}{Pearson product-moment} & Dosage & \parbox[t]{11.5cm}{The Pearson product-moment correlation coefficient between the patient dosage and $\{0,1\}$ indicating if the patient experienced the medical event within 30 days.} \\
Repeat$_{1}$ & Experiment & \parbox[t]{11.5cm}{Number of patients that have medical event in at least two distinct hazard periods and not in their non-hazard periods divided by the number of patients that have at least two distinct hazard periods and have medical event in one hazard period.}\\
Repeat$_{2}$ & Experiment & \parbox[t]{11.5cm}{Number of patients that have medical event in two distinct hazard periods and not in their non-hazard periods divided by the number of times the medical event occurs in the non-hazard periods of patients that have at least two distinct hazard periods.}\\
\end{tabular}
\end{table*}
In this paper we generate association strength attributes by calculating each association measure under three different criteria.  The first criteria considers all drug prescriptions in the database, the second criteria only considers drug prescriptions where the drug is prescribed for the first time in 13 months and the final criteria only considers drug prescriptions where the drug BNF is recorded for the first time in 13 months.  

An existing LOD algorithm for signalling ADRs, known as the Observe to Expected Ratio algorithm \cite{noren2010}, developed a measure of dependency between a drug and medical event known as the $IC_{\Delta}$.  In the paper we also include the $IC_{\Delta}$ and it's lower 95\% confidence value as two additional association strength measures.  So we generate eleven association strength attributes in total.   

\subsubsection{Temporality}
The existing algorithms that detect side effects using LODs have often developed filters that use the temporal information to remove medical events that are associated to the drug of interest but frequently occur before the drug is prescribed.  The Observe to Expected ratio \cite{noren2010} filters medical events that have an $IC_{\Delta}$ on the day of the prescription or the month before the prescription greater than the month after the prescription.  This prompts two temporal attributes $OE_{filt1}$ and $OE_{filt2}$. The $OE_{filt1}$ is one when the $IC_{\Delta}$ for the month before the prescription is greater than the $IC_{\Delta}$ for the month after the prescription and zero otherwise. The $OE_{filt2}$ is one when the $IC_{\Delta}$ on the day of the prescription is greater than the $IC_{\Delta}$ for the month after the prescription and zero otherwise.

Another existing temporality attribute is LEOPARD \cite{Schuemie2011}. LEOPARD takes the perspective of the event and finds how often a drug is prescribed before and after the first occurrence of the event respectively.  If the number of times a drug is prescribed after the first occurrence of the event is significantly greater than the number of times the drug is prescribed before then LEOPARD filters the drug and medical event pair.  The test used for significance is a one sided binomial test.

The final attributes of interest for temporality, the After Before (AB) ratios, are similar to LEOPARD but take the perspective of the drug. The AB ratios calculate how often the medical event occurs after the drug of interest compare to how often the medical event occurs before the drug of interest. It is common for doctors to initially record a less specific medical event and then later in time record a more specific version of the medical event after laboratory results are returned.  As a consequence, many specific medical events that started before the drug was prescribed are only recorded after the prescription, but more general versions of the medical events are recorded before the prescription. This can lead to a high AB ratio for medical events that actually caused the drug to be prescribed.  Rather than calculating the AB ratio for the specific medical event, we decided to calculate the AB ratio for the corresponding more general versions of the medical event, this can be done easily due to medical events being recorded via Read Codes that have a hierarchal structure.  The hierarchal structure means that medical events can be level 1 (very general) to level 5 (very specific), but a level 5 medical event has a corresponding level 1-4 medical event (it's parent, grandparent, great-grandparent, etc...).  Therefore the AB ratios of interest are the AB ratio level 2 that first transforms level 5 to level 3 Read Codes into level 2 ones and then calculates the AB ratio for the level 2 version of the Read Code corresponding to the medical event of interest and the AB ratio level 3 that is similar but only transforms the level 5 to level 4 Read Codes and calculates the AB ratio for the level 3 Read Code version corresponding to the medical event of interest.    

\subsubsection{Specificity}
The specificity measure investigates how specific the association is.  In this paper we consider how specific the medical event is, we investigate if the medical event only occurs for a certain age or gender and we consider if the medical event associated to the drug is also associated to other similar drugs. 

The specificity of the medical event is simply the level of the medical event's Read Code (level 1 to Level 5).  To determine how specific the ages are for patients experiencing the medical event after the drug we calculate the standard deviation of the ages for the patients who have the drug and then experience the medical event and divide this by the standard deviation of the ages for all the patients prescribed the drug. Similarly, to determine if the association tends to only occur for a specific gender we calculate the male/female proportion for patients that take the drug and experience the medical event and divide this by the male/female proportion for all the patents prescribed the drug.
The final values of interest investigate how specific the medical event association is to the drug of interest compared to other drugs in the same BNF family.  A simple measure is to compare the risk ratio of the drug and medical event with the risk ratio of the BNF corresponding to the drug and medical event. 

\subsubsection{Dosage}
One of the criteria that can be calculated using LOD data is the medical event's dependency on the dosage of the drug.  In general, ADRs will be more common in patients prescribed a higher dosage of the drug.  This inspired the following dosage attributes of interest.  The first attribute is the average dosage of the patients experiencing the medical event after the drug divided by the average dosage of the patients prescribed the drug.  If the patients experiencing the medical event after the drug have a higher average dosage than the all the patients prescribed the drug then this indicates that the medical event occurs more often in patients prescribed a high dosage.  Another way of determining if patients prescribed a higher dosage are more likely to experience the medical event is to compare the proportion of patients who are prescribed the highest dosage who also experience the medical event divided by the proportion of patients who are prescribed the lowest dosage who also experience the medical event (if none of the patients prescribed the lowest dosage experience the medical event the denomination applied is 0.95).  The final dosage attributes investigated are the Spearman's rank correlation coefficient and Pearson product-moment correlation coefficient on the set of patient $2-$tuple corresponding to the dosage prescribed and indication if the event occurred, for example $(250,1)$ corresponds to a patient prescribed $250$mg who experienced the event.

\subsubsection{Experimentation}
The experimentation criterion indicates whether the medical event stops and restarts with the drug being stopped and restarted. This attribute only considered patients that have experienced a repeat of the drug with over a year between prescriptions; these patients are referred to as repeat patients.  The hazard period is 30 days after the drug is prescribed and the corresponding non-hazard period is 335 days before the drug prescription.  The Repeat$_{1}$ attribute is calculated as the number of repeat patients that experience the medical event in at least two separate hazard periods and not in the corresponding non-hazard periods divided by the number of repeat patients that experience the medical event in one hazard period and not in the corresponding non-hazard period. The Repeat$_{2}$ attribute is calculated as the number of repeat patients that experience the medical event in at least two separate hazard periods and not in the corresponding non-hazard periods divided by the number of times the medical event occurs in a repeat patients non-hazard period.

\SubSection{Step 2: Feature Selection}
In this study we apply a multivariate filter, the Correlation-based Feature Selection (CFS) algorithm \cite{Hall1999}, as this algorithm is not dependent on a specific classifier.  The CFS algorithm finds the optimal feature subset based on the trade-off between how correlated the class labels are to the feature subset and how intercorrelated the features of the subset are. 

The data used in this study are extracted from The Health Improvement Network database (www.thin-uk.com) and can be found at: http://www.ima.ac.uk/reps. 
\Section{Results}\label{results}
Table \ref{tab:res} shows that the optimal attribute subset to use for ADR discovery is LEOPARD, RD$_{13BNF}$, ABratio Level 3, Gender Ratio and Read Code Level.  The temporal and strength attributes had the greatest correlation with the class labels, whereas 75\% of the dosage attributes has a zero correlation measure. 

\Section{Discussion}
\label{diss}
The results show that the temporal and strength attributes are key for signalling ADRs as these had the highest correlation with the class labels but the specificity attributes Gender Ratio and Read Code level offered potentially new in sight than available via the temporal and strength attributes.  The experiment and dosage attributes investigated in this paper did not offer sufficient additional information than what could be gained from the $RD_{13BNF}$ or the LEOPARD attributes, although there does appear to be some correlation between the class labels and both the Pearson's correlation rank attribute and the Repeats attributes.  

The reason the dosage attributes did not have a greater correlation with the class labels may be due to a limiting factor of comparing different measurement types.  The dosages can be recorded via different measurement types for example `mg', `\%', `mm x cm xcm' or the measure type may be missing.  As it is difficult to determine if $x$ quantity of `mg' is greater than $y$ quantity of '\%', the dosage attributes were calculated only considering prescriptions measured in `mg' (as this was the most popular). Unfortunately this resulted in occasional issues due to `mg' measured prescriptions of some drugs investigated always being the same quantity or many prescriptions of a drug not being included in the dosage attribute calculations.  The experiment attributes were also limited if the drug investigated was rarely repeated.  Furthermore, the experiment attributes may have been biased in this study due to using known ADRs, as if an ADR is known and a patient experiences the ADR after the drug then the doctor is likely to notice this and not prescribed the drug to that patient in the future.  One possible way to overcome this issue would be to use only newly discovered ADRs in the data as the medical records may be more likely to have patients, who at the time unknowingly experienced the ADRs, having a repeat prescription.
\begin{table}
\centering
\caption{The results of the CFS algorithm ordered by the measure of correlation with the class labels.  Attributes not selected by the CFS algorithm have the attribute they are most correlated to listed in the CFS rank column.}
\label{tab:res}
\begin{tabular}{cccc}
Attribute & Class Correlation & CFS Rank \\ \hline
LEOPARD & 0.3238 & 1 \\
OE$_{filt1}$ & 0.2637 & LEOPARD\\ 
OE$_{filt2}$ & 0.2618 & LEOPARD\\
RD$_{13BNF}$ & 0.2347 & 2 \\
RD$_{13d}$ & 0.2248 & LEOPARD \\
RD & 0.2231 & RD$_{13BNF}$ \\
ABratio Lv3 & 0.2231 & 3 \\
ABratio Lv2 & 0.1755 & ABratio Lv3 \\
RR$_{13d}$ & 0.1593 & RD$_{13BNF}$ \\
OR$_{13d}$ & 0.1593 & RD$_{13BNF}$ \\
RR$_{13BNF}$ & 0.1514 & RD$_{13BNF}$ \\
OR$_{13BNF}$ & 0.1514 & RD$_{13BNF}$ \\
RR & 0.1408 & RD$_{13BNF}$ \\
OR & 0.1408 &RD$_{13BNF}$ \\
lowerIC$_{\Delta}$ & 0.135 & RD$_{13BNF}$ \\
Pearson rank & 0.1029 & RD$_{13BNF}$ \\
Gender Ratio & 0.0663 & 4 \\
Repeats$_{1}$ & 0.0651 & LEOPARD \\ 
Repeats$_{2}$ & 0.0651 & LEOPARD \\ 
IC$_{\Delta}$ & 0.0608 &  RD$_{13BNF}$ \\
Read Code Lv & 0.0279 & 5 \\
RR$_{Drug}$/RR$_{BNF}$ & 0 & - \\
Dosage Ratio & 0 & - \\
High Low Ratio & 0 & - \\
Age STDEV & 0 & - \\
Spearman's' rank & 0 & - \\
\end{tabular}
\end{table}

\Section{Conclusion}
\label{con}
In this paper we have applied feature selection to attributes we generated based on the Bradford Hill causality criteria to determine suitable attributes to be used by a general learning algorithm to identify side effects in LODs.  This is the first time suitable attributes for identifying causal relations between prescribed drugs and medical events have been explored and the results now present the opportunity to develop novel learning algorithms.  We have found that the specificity attributes offer additional information for ADR signalling and it would be advantageous to include them into ADR signalling algorithms.  Unfortunately the experiment and dosage attributes were not very correlated with the class labels but this is likely to be due to current limitations.    

Possible future work could focus on developing a way to compare prescriptions with different measurement types so all the prescription data can be used for calculating the dosage attributes or involve developing attributes that cover the remaining Bradford Hill causality criteria (plausibility, coherence, consistency and analogy). 
      
\nocite{ex1,ex2}
\bibliographystyle{latex8}
\bibliography{LitRevRef2_3authors}

\end{document}